\documentclass[preprint,prc,aps]{revtex4}
\usepackage{latexsym}
\everymath={\displaystyle}
\usepackage{rotating}
\usepackage{young}
\def\1{\mbox{l\hspace{-0.53em}1}}
\newcommand{\fr}{\frac}

\begin{document}
\title{Matrix elements of SU(6) generators for baryons with arbitrary $N_c$
quarks in mixed symmetric states $[N_c-1,1]$}

\author{N. Matagne$^a$\footnote{e-mail address: Nicolas.Matagne@theo.physik.uni-giessen.de}}

\author{Fl. Stancu$^b$\footnote{e-mail address: fstancu@ulg.ac.be
}}
\affiliation{
$^a$ Institut f\"ur Theoretische Physik, Universit\"at Gie\ss en, D-35392 Gie\ss en, Germany \\
$^b$ University of Li\`ege, Institute of Physics B5, Sart Tilman,
B-4000 Li\`ege 1, Belgium}

\date{\today}

\begin{abstract}
We derive general analytic formulae for the matrix elements of the SU(6) 
generators for mixed symmetric $[N_c-1,1]$ spin-flavor states with an arbitrary
number $N_c$ of quarks. They are relevant for baryon spectroscopy in 
the $1/N_c$ expansion method applied to light baryons and can
be used to study excited states. 
In this way  previous work on non-strange baryons can be extended 
to both non-strange and strange baryons.
\end{abstract}

\maketitle

\section{Introduction}

In the $1/N_c$ expansion method \cite{HOOFT,WITTEN} the ground state baryons 
have an approximate SU(2$N_f$) symmetry when the number of colors $N_c$
is large but finite. This stems from the property that when
 $N_c \rightarrow \infty$ the SU(2$N_f$) is an exact contracted symmetry
\cite{Gervais:1983wq,DM} 
and in that limit the baryons are degenerate.  
At large but finite $N_c$ the mass splitting starts at order $1/N_c$  
for the ground or excited symmetric states and at order $N^0_c$
for mixed symmetric states.

Here we consider light
baryons with $N_f = 3$, thus we deal with SU(6) symmetry. In this case
the building blocks of the mass operator are the generators of SU(6).
For the excited states the symmetry is extended to SU(6) $\times$ SO(3). 
Therefore the generators of SO(3) also appear in the mass formula.

The study of excited states of symmetric orbital symmetry 
is straightforward.  
However the study of excited states of mixed orbital symmetry, or equivalently 
mixed spin-flavor symmetry, presented so far some difficulty related to the 
fact the matrix elements of SU(6) generators between mixed symmetric 
$[N_c-1,1]$ spin-flavor states were unknown. Accordingly, a method based on 
the separation of a system of $N_c$ quarks into a symmetric ``core'' of 
$N_c-1$ quarks and an excited quark was proposed \cite{CCGL} and applied to
the $[{\bf 70},1^-]$ and $[{\bf 70},\ell^+]$ ($\ell = 0, 2$) multiplets, for
 $N_f = 2$ \cite{CCGL,Matagne:2005plb} and $N_f=3$ \cite{Goity:2002pu,Matagne:2006zf}. 
The orbital-spin-flavor wave function describing such a decoupled system
is not totally symmetric, as it should be. Its approximate form
is explained at large in Ref. \cite{Matagne:2008fw}.
In addition, to match the decoupling, in Ref. \cite{CCGL}
each generator of SU($2N_f$) was written as a sum of two
terms, one acting on the core and the other on the excited quark. 
As a consequence, the number of linearly independent operators 
appearing in the mass formula increases 
and the number of coefficients  
to be determined  generally becomes larger or nearly as large as
the number of the experimental data
available. 
For example, for the $[{\bf 70},1^-]$ multiplet with $N_f = 3$ 
one has at least 15 linearly independent
operators up to order $1/N_c$  included \cite{CCGL} and 7 known masses. 
Then one must select the most dominant operators, which is a very
difficult task,  not free from ambiguities \cite{Goity:2002pu}.

In Ref. \cite{Matagne:2006dj} we have proposed a new approach where 
the separation of the system into a symmetric core of $N_c-1$ quarks 
and an excited quark can be avoided. The approach was restricted to
$N_f$ = 2. In this way the number of linearly 
independent operators was substantially reduced. In addition the
method has important physical consequences. We have shown for example  
that the term containing the isospin-isospin interaction in the mass 
formula of $\Delta$, neglected in all previous applications, becomes
as dominant as the pure spin-spin term in $N$. 
To obtain such results the
matrix elements of the generators of SU(4) were needed. 
General analytic expression were
available from nuclear physics studies \cite{HP}. It was easy enough
to adjust them to a system of $N_c$ quarks. 

Let us explain the previous situation in detail. In the scheme based on the  
separation of the system into a symmetric core of $N_c-1$ quarks
and an excited quark the SU(2)-isospin Casimir operator was written as
$T^2 = T^2_c + 2 t \cdot T_c + 3/4$,
where the lower index $c$ refers to the core, and decomposed into 
three independent pieces, corresponding to the terms in the above decomposition. 
In SU(4) $T^2_c$ and $S^2_c$ have identical matrix elements because
the spin and isospin states of a symmetric core are identical, so that
$T^2_c$ can be neglected. But $t \cdot T_c$ has different matrix
elements from $s \cdot S_c$ as one can clearly see from Table II
of Ref. \cite{CCGL}. Then in the decoupling scheme
the isospin can be introduced only through $t \cdot T_c$.
In Ref. \cite{Matagne:2008fw} Table VI
we have shown that the introduction of the operators 
$\frac{1}{N_c}t \cdot T_c$ together with  $\frac{1}{N_c}S^2_c$ 
and $\frac{1}{N_c}s \cdot S_c$ separately deteriorates the fit.
This may explain why $\frac{1}{N_c}t \cdot T_c$ has been avoided in previous 
numerical fits both in SU(4) \cite{CCGL} and in SU(6) 
\cite{Goity:2002pu}. We avoided it as well  \cite{Matagne:2005plb}
in line with our predecessors.

To extend the application of the method of Ref. \cite{Matagne:2006dj} from non-strange to both
non-strange and strange baryons one 
needs to know the matrix elements of the generators of SU(6). 
In this work we derive these matrix elements.

We recall that the group SU(6) has 35 generators ${S^i,T^a,G^{ia}}$
with $i = 1,2,3$ and $a = 1,2,\ldots,8$ where $S^i$ are the generators 
of the spin subgroup SU(2) and $T^a$ the generators of the flavor 
subgroup SU(3). The group algebra is
\begin{eqnarray}\label{ALGEBRA}
&[S^i,S^j]  =  i \varepsilon^{ijk} S^k,
~~~~~[T^a,T^b]  =  i f^{abc} T^c, \nonumber \\
&[S^i,G^{ja}]  =  i \varepsilon^{ijk} G^{ka},
~~~~~[T^a,G^{jb}]  =  i f^{abc} G^{ic}, \nonumber \\
&[G^{ia},G^{jb}] = \fr{i}{4} \delta^{ij} f^{abc} T^c
+\fr{i}{2} \varepsilon^{ijk}\left(\fr{1}{3}\delta^{ab} S^k 
+d^{abc} G^{kc}\right),
\end{eqnarray}
by which the normalization of the generators is fixed.

We redefine the generators forming the algebra (\ref{ALGEBRA}) as 
\begin{equation} \label{normes}
E^i =\frac{ S^i}{\sqrt{3}};~~~ E^a = \frac{T^a}{\sqrt{2}}; ~~~E^{ia} = \sqrt{2}
G^{ia} .
\end{equation}
Note that the generic name  for every generator will also be $E^{ia}$ \cite{HP}.
Specifications will be made whenever necessary.
Here we search for the matrix elements of  $S^i$, $T^a$ 
and $G^{ia}$ between SU(6) states of symmetry $[N_c-1,1]$.
As we shall see below, the matrix elements of $S^i$ and $T^a$
are straightforward. The remaining problem is to derive
the matrix elements of $G^{ia}$.

The SU(6) generators are the components of an irreducible tensor operator 
which transform according to the adjoint representation $[21^4]$, equivalent to 
${\bf 35}$, in dimensional notation.   There are several ways to
calculate the matrix elements of the SU(6) generators. In the standard
group theory 
the matrix elements of any irreducible tensor can be expressed in
terms of a generalized Wigner-Eckart theorem which is a factorization
theorem, involving the product between a reduced matrix element and 
a Clebsch-Gordan (CG) coefficient.  The CG coefficient of SU(6) factorizes into
CG coefficients of SU(2), SU(3) and an isoscalar factor of SU(6), see Eq. 
(\ref{GEN}).  The latter is the quantity we derive here. 

  In the 60'ties the literature provided a few examples of isoscalar factors
needed in particle physics, thus for $N_c = 3$.
Cook and Murtanza \cite{CM65} considered the full
CG series of the direct products 
${\bf 35} \times {\bf 35}$,  ${\bf 56} \times {\bf 35}$ and  
${\bf 56} \times \overline {\bf 56}$. 
Carter, Coyne and Meshkov \cite{CCM} derived the isoscalar factors for 
${\bf 56 \times 35 \rightarrow 56}$ independently from  Sch\"ulke 
\cite{SCHULKE} who also
calculated the isoscalar factors for  ${\bf 35 \times 35 \rightarrow 35}$
like Cook and Murtanza. Moreover
Carter and Coyne \cite{CC} derived the isoscalar factors of the whole CG series
${\bf 35} \times {\bf 70} = {\bf 20}+{\bf 56}+2 \times {\bf 70}+{\bf 540}
+{\bf 560}+{\bf 1134}$. In Ref. \cite{Matagne:2006xx}
we have obtained analytic formulae for isoscalar factors of arbitrary $N_c$
which for $N_c = 3$ correspond to ${\bf 56 \times 35 \rightarrow 56}$.
Up to a phase we have found full agreement with Refs. \cite{CM65,CCM,SCHULKE}.  
In the present case, by setting $N_c = 3$ in our formulae, we
could, in principle, compare the results with either column $70_{\mathrm{I}}$
or column $70_{\mathrm{II}}$ of Ref. \cite{CC},   
for ${\bf 35} \times {\bf 70} \rightarrow {\bf 70}$. 
In fact, following our definition,  one does not need to compute the isoscalar 
factors of both  $70_{\mathrm{I}}$
and $70_{\mathrm{II}}$ to derive the matrix elements of the generators.
However a multiplicity 2 problem can appear in the direct product of two SU(3) 
irreducible representations (see Eq. (\ref{PROD}) below). For $N_c = 3$, this 
is the case for the product $(8\times 8) \to 8$. Here we shall use the label 
$\rho$ \cite{Matagne:2006xx} to distinguish between the two representations
when the multiplicity is 2. In Ref. \cite{CC},  the authors
follow the notation of \cite{DESWART} by using the label $S$ for the symmetric
product and $A$ for antisymmetric product corresponding respectively to $\rho =2$ and 
$\rho = 1$ in our notation.
For consistency with previous work 
\cite{Matagne:2006xx}, we follow 
the definition of Ref. \cite{HP}, (see Eq. (\ref{REDUCED}) below), which 
simplifies the problem, in the sense that we obtain vanishing SU(6) isoscalar 
factors for the products $(8^{4,2} \times 8^1)_{\mathrm{S}}$ while Carter 
and Coyne obtain non-vanishing values for the isoscalar factors for these products. 
Thus the comparison is impossible.
Such ambiguities are typical for all groups, including the permutation
group, whenever the multiplicity in the CG series is larger than one
\cite{ISOSC}. As mentioned above, for unitary groups, following Gell-Mann, 
it is customary to introduce the 
symmetric $S$ or $D$ coupling and  antisymmetric $A$ or $F$ coupling. 
The choice is based on convenience anyhow \cite{DON}.

The paper is organized as follows. In the next section we introduce 
the SU(6) basis states. In Sec. III we remind the generalized 
Wigner-Eckart theorem which allows a factorization of the
matrix elements into  Clebsch-Gordan coefficients and some
specific isoscalar factors. In Sec. IV we derive the unknown 
isoscalar factors. In Sec. V we discuss possible physical applications
to the mass spectrum and in the last section we summarize our results.   

\section{The wave function}\label{WF}

We deal with a system of $N_c$ quarks having one unit of
orbital excitation. Therefore the orbital ($O$) wave function  must have 
a mixed symmetry $[N_c-1,1]$, which describes the lowest excitations in
a baryon.
The fact that this is the lowest excitation with $L$ = 1 is well known in 
group theory and has been extensively applied to
nuclear shell model, see {\it e.g.} \cite{TALMI} or
\cite{HECHT64}. 
Moreover the $N_c-1$ independent basis states of the $[N_c-1,1]$ irrep  
written in the 
Young-Yamanouchi basis, see below, is equivalent to a basis
written in terms of $N_c-1$  internal Jacobi coordinates, thus
the center of mass motion is automatically removed.
An example for four quarks can be found in Ref. \cite{Stancu:1998sm}. 

The colour wave function being antisymmetric, the
orbital-spin-flavor wave part must be symmetric. Then the spin-flavor ($FS$)
part must have the same symmetry as the orbital part in order to obtain a 
totally symmetric
state in the orbital-spin-flavor space. 
The general form of such a wave function is \cite{book}
\begin{equation}
\label{EWF}
|[N_c] \rangle = {\frac{1}{\sqrt{d_{[N_c-1,1]}}}}
\sum_{Y} |[N_c-1,1] Y \rangle_{O}  |[N_c-1,1] Y \rangle_{FS},
\end{equation}    
where $d_{[N_c-1,1]} = N_c - 1$ is the dimension of the representation 
$[N_c-1,1]$ of the permutation group $S_{N_c}$ and $Y$ is a symbol for a
Young tableau (Yamanouchi symbol). 
The sum is performed over all possible standard Young tableaux. 
In each term the first basis vector represents the orbital  
space  and the second the spin-flavor space.
In this sum there is only one $Y$ (the normal Young tableau) 
where the last particle is in 
the second row and $N_c - 2 $ terms where the last particle
is in the first row.
The explicit form of the orbital part is not needed.

More precisely, we write $Y = (pqy)$ where $p$ is the row of
the $N_c$-th particle, $q$ the row of the  $(N_c-1)$-th particle and $y$
is the Young tableau of the remaining particles.
Let us denote by $p$, $p'$ and $p''$ the position of
the last particle in the spin-flavor, spin and flavor Young tableaux
respectively. They are indicated by crosses in the example 
given by Eqs. (\ref{decupletch52})-(\ref{singletch52}) below.
Similarly for the $(N_c-1)$-th particle we have $q$, $q'$ and $q''$ and for the
rest $y$, $y'$ and $y''$.
We need now to decompose the spin-flavor wave function into its spin and flavor
parts separately. 
For this purpose we use the Clebsch-Gordan (CG) coefficients of $S_{N_c}$, 
denoted by $S([f']p'q'y' [f'']p''q''y'' | [f]pqy )$ 
and their factorization property \cite{book}.    
Denoting by $K([f']p'[f'']p''|[f]p)$ the isoscalar factors of $S_{N_c}$ 
we have \cite{Matagne:2008fw}
\begin{equation}\label{racah} 
S([f']p'q'y' [f'']p''q''y'' | [f]pqy ) =
K([f']p'[f'']p''|[f]p) S([f'_{p'}]q'y' [f''_{p''}]q''y'' | [f_p]qy ),
\end{equation}
where the second factor in the right-hand side is a CG coefficient of
$S_{N_c-1}$ containing the 
partitions $[f'_{p'}]$, $[f''_{p''}]$ and $[f_p]$ obtained after 
the removal of the $N_c$-th quark.

Using the above property we can write the spin-flavor part of the wave 
function as
\begin{eqnarray}\label{fs}
|[N_c-1,1]p;(\lambda\mu) Y I I_3;S S_3 \rangle = 
\sum_{p' p''} K([f']p'[f'']p''|[N_c-1,1]p) 
|S S_3;p' \rangle|(\lambda\mu) Y I I_3;p'' \rangle   
\end{eqnarray}
where the spin part is
\begin{equation}\label{spin}
|S S_3; p' \rangle = \sum_{m_1,m_2}
 \left(\begin{array}{cc|c}
	S_c    &    \frac{1}{2}   & S   \\
	m_1  &         m_2        & S_3
      \end{array}\right)
      |S_cm_1 \rangle |1/2m_2 \rangle,
\end{equation}
with $S_c = S - 1/2$ if $p' = 1$    and $S_c = S + 1/2$ if   $p' = 2$ and
the flavor part is 
\begin{eqnarray}
\lefteqn{|(\lambda\mu) Y I I_3,p'' \rangle =} \nonumber \\ & &
\sum_{Y_c,I_c,I_{c_3},\atop yii_3} \left(\begin{array}{cc||c}
	(\lambda_c\mu_c)    &    (10)   &  (\lambda\mu)  \\
	Y_cI_c &         yi        & YI
      \end{array}\right)
\left(\begin{array}{cc|c}
	I_c    &    i   & I   \\
	I_{c_3}  &   i_3              & I_3
      \end{array}\right)
|(\lambda_c\mu_c)Y_cI_cI_{c_3}\rangle |(10)yii_3 \rangle,
\end{eqnarray}
with $(\lambda_c,\mu_c) = (\lambda-1,\mu)$ for $p'' = 1$, $(\lambda_c,\mu_c) = (\lambda+1,\mu-1)$ for $p'' = 2$ and $(\lambda_c,\mu_c) = (\lambda,\mu+1)$ for $p'' = 3$.
Each SU(3) irreducible representation carries the label $(\lambda\mu)$. 

Let us illustrate Eq. (\ref{fs}) in terms of Young tableaux by taking 
the $N_c = 7$ and $p = 2$. We have the following cases

\begin{eqnarray}
\label{decupletch52}
\raisebox{-9.0pt}{\mbox{\begin{Young}
 & & & & & \cr
$\times$ \cr
\end{Young}}}\
& = &
K([43]2[52]1|[61]2)\! \! \!
\raisebox{-9.0pt}{\mbox{
\begin{Young}
& & & \cr
& & $\times$\cr
\end{Young}}} \ \times \! \! \! \! \!
\raisebox{-9.0pt}{\mbox{
\begin{Young}
& & & & $\times$\cr
& \cr
\end{Young}}}\ ,
\\ \nonumber
\\
\label{octetch54}
\raisebox{-9.0pt}{\mbox{\begin{Young}
 & & & & & \cr
$\times$ \cr
\end{Young}}}\
& = &
K([52]1[43]2|[61]2)\! \! \!
\raisebox{-9.0pt}{\mbox{
\begin{Young}
& & & & $\times$\cr
& \cr
\end{Young}}} \ \times \! \! \! \! \!
\raisebox{-9.0pt}{\mbox{
\begin{Young}
& & & \cr
& & $\times$\cr
\end{Young}}}\ ,
\\ \nonumber
\\
\label{octetch52}
\raisebox{-9.0pt}{\mbox{\begin{Young}
 & & & & & \cr
$\times$ \cr
\end{Young}}}\
&=& K([43]1[43]1|[61]2)\! \! \!
\raisebox{-9.0pt}{\mbox{
\begin{Young}
& & & $\times$\cr
& & \cr
\end{Young}}} \ \times \! \! \! \! \!
\raisebox{-9.0pt}{\mbox{
\begin{Young}
& & & $\times$\cr
& & \cr
\end{Young}}} \nonumber \\
& & + \ K([43]2[43]2|[61]2)\! \! \!
\raisebox{-9.0pt}{\mbox{
\begin{Young}
& & & \cr
& & $\times$\cr
\end{Young}}} \ \times \! \! \! \! \!
\raisebox{-9.0pt}{\mbox{
\begin{Young}
& & & \cr
& & $\times$\cr
\end{Young}}}\ ,
\\ \nonumber
\\
\label{singletch52}
\raisebox{-9.0pt}{\mbox{\begin{Young}
 & & & & & \cr
$\times$ \cr
\end{Young}}}\
&=& K([43]1[331]3|[61]2)\! \! \!
\raisebox{-9.0pt}{\mbox{
\begin{Young}
& & & $\times$\cr
& & \cr
\end{Young}}} \ \times \! \! \! \! \!
\raisebox{-15pt}{\mbox{
\begin{Young}
& & \cr
& & \cr
$\times$ \cr
\end{Young}}}\ .
\end{eqnarray} 
When $N_c = 3$ the above spin-flavor states correspond to
$^{2}10$, $^{4}8$, $^{2}8$ and $^{2}1$ multiplets. 
For the purpose of the present study we actually need only the 
case $p$ = 2. In this case the isoscalar factors $K([f']p'[f'']p''|[N_c-1,1]p)$ 
for arbitrary $N_c$ have the following algebraic form \cite{Matagne:2005plb}

\begin{eqnarray}
K\left(\left[\frac{N_c+1}{2},\frac{N_c-1}{2}\right]2\left[\frac{N_c+3}{2},\frac{N_c-3}{2}\right]1|[N_c-1,1]2\right) & = & 1,\nonumber \\
K\left(\left[\frac{N_c+3}{2},\frac{N_c-3}{2}\right]1\left[\frac{N_c+1}{2},\frac{N_c-1}{2}\right]1|[N_c-1,1]2\right) & = & 1,\nonumber \\
 K\left(\left[\frac{N_c+1}{2},\frac{N_c-1}{2}\right]1\left[\frac{N_c+1}{2},\frac{N_c-1}{2}\right]1|[N_c-1,1]2\right) & = & -\sqrt{\frac{3(N_c-1)}{4N_c}}, \nonumber \\
K\left(\left[\frac{N_c+1}{2},\frac{N_c-1}{2}\right]2\left[\frac{N_c+1}{2},\frac{N_c-1}{2}\right]2|[N_c-1,1]2\right) & = & \sqrt{\frac{N_c+3}{4N_c}},\nonumber \\
K\left(\left[\frac{N_c+1}{2},\frac{N_c-1}{2}\right]1\left[\frac{N_c-1}{2},\frac{N_c-1}{2},1\right]3|[N_c-1,1]2\right) & = & 1.
\end{eqnarray}
For completeness we mention that the isoscalar factors for $p = 1$ and  non-strange states 
can be found in Ref. \cite{Matagne:2008fw}.

\section{Matrix elements of SU(6) generators: 
the generalized Wigner-Eckart theorem}

For the spin  $S^i$ and the flavor  $T^a$ operators the matrix elements
can be obtained from the Wigner-Eckart theorem in a similar manner
as for symmetric $N_c$ states \cite{Matagne:2006xx}. As already mentioned,
the SU(6) generators are the components of an irreducible tensor operator 
which transform according to the adjoint representation $[21^4]$, equivalent to 
${\bf 35}$, in dimensional notation.   
The matrix elements of any irreducible tensor can be expressed in
terms of a generalized Wigner-Eckart theorem which is a factorization
theorem, involving the product between a reduced matrix element and 
a Clebsch-Gordan (CG) coefficient. 
The case SU(4) $\supset$ SU(2) $\times$ SU(2)
has been worked out by Hecht and Pang \cite{HP} in a 
general form needed for  applications to
nuclear physics. 

By using the generalized Wigner-Eckart theorem, in Ref. \cite{Matagne:2006xx} 
we have derived explicit formulas for the matrix elements of SU(6)
generators for symmetric states of $N_c$ quarks of  
partition $[N_c]$. Here we use a different procedure to obtain 
the matrix elements of SU(6)  generators between mixed symmetric states
$[N_c-1,1]$.  When $N_c = 3$ they correspond to the representation $[{\bf 70}]$.

By analogy to SU(4)  \cite{HP} one can write
the  matrix elements of every SU(6) generator $E^{ia}$ as
\begin{eqnarray}\label{GEN}
\lefteqn{\langle [N_c-1,1](\lambda' \mu') Y' I' I'_3 S' S'_3 | E^{ia} |
[N_c-1,1](\lambda \mu) Y I I_3 S S_3 \rangle =}\nonumber \\ & & \sqrt{C^{[N_c-1,1]}(\mathrm{SU(6)})} 
   \left(\begin{array}{cc|c}
	    S   &    S^i   & S'   \\
	    S_3  &   S^i_3   & S'_3
  \end{array}\right)
     \left(\begin{array}{cc|c}
	I   &   I^a   & I'   \\
	I_3 &   I^a_{3}   & I'_3
   \end{array}\right)  \nonumber \\
& & \times       \sum_{\rho = 1,2}
 \left(\begin{array}{cc||c}
	(\lambda \mu)    &  (\lambda^a\mu^a)   &   (\lambda' \mu')\\
	Y I   &  Y^a I^a  &  Y' I'
      \end{array}\right)_{\rho}
\left(\begin{array}{cc||c}
	[N_c-1,1]    &  [21^4]   & [N_c-1,1]   \\
	(\lambda \mu) S  &  (\lambda^a\mu^a) S^i  &  (\lambda' \mu') S'
      \end{array}\right)_{\rho} , 
   \end{eqnarray}
where $C^{[N_c-1,1]}(\mathrm{SU(6)}) = N_c(5 N_c+18)/12$ is the SU(6)
Casimir operator associated to 
the irreducible representation $[N_c-1,1]$, followed by Clebsch-Gordan
coefficients of SU(2)-spin and SU(2)-isospin. The sum over $\rho$ is over
terms containing products of isoscalar factors of SU(3) and SU(6) respectively.
In particular, $T^a$ is an SU(3) irreducible tensor operator of components
$T^{(11)}_{Y^aI^a}$, \emph{i.e.} $a$ corresponds to $(\lambda^a\mu^a)$. 
It is a scalar in SU(2) so that the index $i$ from $E^{ia}$
is no more necessary. The generators $S^i$
form a rank 1 tensor in SU(2) which is a scalar in SU(3), so the index $i$ suffices.
Although we use the same symbol for the operator $S^i$ and its quantum numbers 
we hope that no confusion is created.
Thus, for the generators $S^i$ and $T^a$, which are elements of the $su$(2) and
$su$(3) subalgebras of (\ref{ALGEBRA}), the above expression
simplifies considerably. In particular, as  $S^i$  acts only on the
spin part of the wave function, we apply the usual
Wigner-Eckart theorem for SU(2) to get
\begin{eqnarray}\label{SPIN}
\lefteqn{\langle [N_c-1,1](\lambda'\mu') Y' I' I'_3; S' S'_3 |S^i|
[N_c-1,1](\lambda \mu) Y I I_3; S S_3 \rangle =} \nonumber \\ & &  \delta_{SS'}\delta_{\lambda \lambda'} \delta_{\mu\mu'} \delta_{YY'} \delta_{II'} \delta_{I_3I_3'}
 \sqrt{C(\mathrm{SU(2)})} \left(\begin{array}{cc|c}
	S   &    1  &  S'   \\
	S_3 &    i  &  S'_3
      \end{array}\right),
   \end{eqnarray}
with $C(\mathrm{SU(2)}) = S(S+1)$.  
Similarly, we use the Wigner-Eckart theorem for $T^a$ which is a generator of 
the subgroup SU(3)
\begin{eqnarray}\label{FLAVOR}
\lefteqn{\langle [N_c-1,1](\lambda'\mu') Y' I' I'_3; S' S'_3 |T^a|
[N_c-1,1](\lambda \mu) Y I I_3; S S_3 \rangle =} \nonumber \\ & &
\delta_{SS'} \delta_{S_3S'_3}\delta_{\lambda \lambda'} \delta_{\mu\mu'}
\sum_{\rho = 1,2}
\langle (\lambda'\mu') || T^{(11)} || (\lambda \mu) \rangle_{\rho}
  \left(\begin{array}{cc|c}
	(\lambda \mu)    &  (11)   &   (\lambda'\mu')\\
	YII_3   &  Y^aI^aI^a_{3}  &  Y' I' I'_3
      \end{array}\right)_{\rho},
   \end{eqnarray}
where the reduced matrix element is defined as  \cite{HECHT} 
\begin{eqnarray}\label{REDUCED}
\langle (\lambda \mu) || T^{(11)} || (\lambda \mu) \rangle_{\rho} = \left\{
\begin{array}{cc}
\sqrt{C(\mathrm{SU(3)})}      & \mathrm{for}\ \rho = 1 \\
0 & \mathrm{for}\ \rho = 2 \\
\end{array}\right.
,\end{eqnarray}   
in terms of the eigenvalue of the Casimir operator       
$C(\mathrm{SU(3)}) = \frac{1}{3} g_{\lambda \mu}$ where
\begin{equation}\label{CSU3}
g_{\lambda\mu}= {\lambda}^2+{\mu}^2+\lambda\mu+3\lambda+3\mu.
\end{equation}
Note that the presence of the index $\rho$ has the same origin as in Eq. 
(\ref{GEN}), namely it reflects the 
multiplicity problem appearing in the direct product of SU(3) irreducible representations 
\begin{eqnarray}\label{PROD}
\lefteqn{(\lambda \mu) \times (11)  =  (\lambda+1, \mu+1)+ (\lambda+2, \mu-1) +
(\lambda \mu)_1 + (\lambda \mu)_2}  \nonumber \\
& & + \, (\lambda-1, \mu+2) + (\lambda-2, \mu+1)
+ (\lambda+1, \mu-2)+ (\lambda-1, \mu-1).
\end{eqnarray}
 Each SU(3) CG coefficient factorizes 
into an SU(2)-isospin CG coefficient and an SU(3) isoscalar factor
\cite{DESWART}
\begin{equation}\label{CGSU3}
\left(\begin{array}{cc|c}
	(\lambda \mu)    &  (11)   &   (\lambda'\mu')\\
	YII_3   &  Y^aI^aI^a_{3}  &  Y' I' I'_3
      \end{array}\right)_{\rho} =
\left(\begin{array}{cc|c}
	I   &    1  &  I'   \\
	I_3 &    I^a_3  &  I'_3
     \end{array}\right)
 \left(\begin{array}{cc||c}
	(\lambda \mu)    &  (11)   &   (\lambda'\mu')\\
 	 YI   &  Y^aI^a  &  Y' I'
      \end{array}\right)_{\rho}.
 \end{equation}   
The analytic expression of the isoscalar factors
can be found in Table 4 of Ref. \cite{HECHT}.

Therefore the basic problem is to determine the matrix elements
of $G^{ia}$. The procedure is described in the next section.

\section{SU(6) isoscalar factors}\label{IF}

\subsection{The general procedure}

A convenient way to derive the matrix elements of $G^{ia}$
is by decoupling the $N_c$-th quark from the system of $N_c$ 
quarks in a mixed symmetric state $[N_c-1,1]$. Let us denote by $G^{ia}$, 
$G^{ia}_c$ and $g^{ia}$
the generators of the total, of the $N_c-1$ system and of the
decoupled quark. Then one has
\begin{equation}\label{DECOUPLE}
G^{ia} = G^{ia}_c + g^{ia}.
\end{equation} 
As mentioned before the last quark can be either in the row $p$ = 1
or in $p$ = 2. Now we use the important observation that the matrix
elements of the generators are independent of the choice of $p$
as used, for example, in  Appendix A of Ref. \cite{Matagne:2008fw}.
The explanation lies in Weyl's duality theorem according to which 
the basis vectors introduced in Sec.~\ref{WF} form invariant subspaces
both for the permutation group and the SU(6) group \cite{book}.
It is therefore useful to take  $p$ = 2 because in that case
the system of $N_c-1$ quarks is in a symmetric  $[N_c-1]$ state
for which the matrix elements are already known from 
Ref. \cite{Matagne:2006xx} where it is enough to replace $N_c$
by $N_c-1$.  The matrix elements of $g^{ia}$ are the trivial case
of symmetric states with $N_c$ = 1.
In a short notation we therefore have
\begin{equation}\label{Gia_p2}
\langle G^{ia} \rangle = \langle G^{ia}_c \rangle_{p}
 + \langle g^{ia} \rangle_{p} 
 \end{equation}
irrespective of the value $p = 1$ or 2.

After lengthy calculations we obtain the following expression for the 
matrix elements in the right-hand side of (\ref{Gia_p2})
\begin{eqnarray}\label{CORE}
 \lefteqn{\langle [N_c-1,1]p;(\lambda'\mu')Y'I'I_3';S'm_s'|G_c^{ja}|[N_c-1,1]p;(\lambda\mu)YII_3;Sm_s\rangle =}\nonumber\\
&  & (-1)^{1/2-S}\sqrt{(2S+1)(2S'_c+1)}\sqrt{\frac{C^{[f]}(SU(6))}{2}}
\left(\begin{array}{cc|c}
 S & 1& S'\\
m_s & j & m'_s
\end{array}\right)
\left(\begin{array}{cc|c}
 I & I^a& I'\\
I_3 & I^a_3 & I'_3
\end{array}\right) \nonumber \\
& &\times \sum_{p',p'',q',q''} (-1)^{S'_c}(-1)^{\lambda-\lambda_c+\lambda'-\lambda'_c}(-1)^{\mu-\mu_c+\mu'-\mu'_c}K([f']p'[f'']p''|[N_c-1,1]p)\nonumber \\ 
& &\times K([f']q'[f'']q''|[N_c-1,1]p) 
\left\{\begin{array}{ccc}
       S & 1 & S' \\
	S'_c & 1/2 & S_c
      \end{array}\right\}
 \sum_{\rho,\rho_c=1,2} \langle (\lambda\mu)YI;(11)Y^aI^a||(\lambda'\mu')Y'I'\rangle_\rho \nonumber \\
&  & \times U((10)(\lambda_c\mu_c)(\lambda'\mu')(11);(\lambda\mu)_\rho(\lambda'_c\mu'_c)_{\rho_c})
\left(\begin{array}{cc||c}
       [f_c] & [21^4] & [f_c] \\
	(\lambda_c\mu_c)S_c & (11)1 & (\lambda'_c\mu'_c)S'_c
      \end{array}\right)_{\rho_c}
\end{eqnarray}
which contains a summation over the indices $\rho$ and $\rho_c$ 
related to the total system of $N_c$ quarks and to the core formed
of $N_c-1$ quarks. One has $[f] =[N_c-2,1]$ for $p=1$ and $[f]=[N_c-1]$ for $p=2$. The SU(3) Racah coefficients $U$ appear due to the 
recoupling of the last quark. 
The matrix elements of the separated quark are
simpler, as expected  
\begin{eqnarray}\label{SEPARATE}
 \lefteqn{\langle [N_c-1,1]p;(\lambda'\mu')Y'I'I_3';S'm_s'|g^{ja}|[N_c-1,1]p;(\lambda\mu)YII_3;Sm_s\rangle =}\nonumber\\
& &(-1)^{S'-1/2}\sqrt{2(2S+1)}\left(\begin{array}{cc|c}
 S & 1& S'\\
m_s & j & m'_s
\end{array}\right)
\left(\begin{array}{cc|c}
 I & I^a& I'\\
I_3 & I^a_3 & I'_3
\end{array}\right) \nonumber \\ &  & \times \sum_{p',p'',q',q''} (-1)^{S_c}K([f']p'[f'']p''|[N_c-1,1]p) K([f']q'[f'']q''|[N_c-1,1]p)
\left\{\begin{array}{ccc}
       S & 1 & S' \\
	1/2 & S_c & 1/2
      \end{array}\right\} \nonumber \\
& & \times\sum_{\rho=1,2} \langle (\lambda\mu)YI;(11)Y^aI^a||(\lambda'\mu')Y'I'\rangle_\rho  U((\lambda_c\mu_c)(10)(\lambda'\mu')(11);(\lambda\mu)(10))_\rho
\end{eqnarray}
They all contain the isoscalar factors $K$ given in the previous section.
Inserting them in the above expressions, together with the isoscalar 
factors $\langle(\lambda\mu)YI;(11)Y^aI^a||(\lambda'\mu')Y'I'\rangle_\rho$
and the SU(3) Racah coefficients $U$ and performing the sums 
in (\ref{CORE}) and (\ref{SEPARATE})
we can obtain the matrix elements described in the next section.

As just mentioned above, for the calculations of the $G_c^{ia}$ and $g^{ia}$ 
matrix elements one needs to derive some SU(3) Racah coefficients. For that 
purpose, we follow the method described by Hecht \cite{HECHT}. We have 
obtained the following formulas to be used in Eqs. (\ref{CORE}) and 
(\ref{SEPARATE})
\begin{eqnarray}
 \lefteqn{\langle(\lambda_c'\mu_c')Y'_cI'_c;(10)yi||(\lambda'\mu')Y'I'\rangle\; U((10)(\lambda_c\mu_c)(\lambda'\mu')(11);(\lambda\mu)_\rho(\lambda'_c\mu'_c)_{\rho_c})=}\nonumber \\
& & \sum_{Y_c,Y_a,Y,\atop I_c,I_a,I} (-1)^{\lambda_c-\lambda+\mu-\mu_c} (-1)^{\lambda'_c-\lambda'+\mu'-\mu'_c}(-1)^{i+I_a+I+I'_c}\sqrt{(2I+1)(2I'_c+1)}  \left\{
\begin{array}{ccc}
i & I_c & I\\
I_a & I'& I'_c
\end{array}\right\} \nonumber \\  & & \times\langle(\lambda_c\mu_c)Y_cI_c;(11)Y_aI_a||(\lambda'_c\mu'_c)Y'_cI'_c\rangle_{\rho_c}
 \langle(\lambda\mu)YI;(11)Y_aI_a||(\lambda'\mu')Y'I'\rangle_{\rho}
\nonumber \\  & & \times
\langle(\lambda_c\mu_c)Y_cI_c;(10)yi||(\lambda\mu)YI\rangle,
\end{eqnarray}
and
\begin{eqnarray}
 \lefteqn{\langle(\lambda_c\mu_c)Y_cI_c;(10)y'i'||(\lambda'\mu')Y'I'\rangle\; U((\lambda_c\mu_c)(10)(\lambda'\mu')(11);(\lambda\mu)_\rho(10))=} \nonumber \\ & &
\sum_{y,Y_a,Y, \atop i, I_a, I}(-1)^{I_c+i+I'+I_a}\sqrt{(2I+1)(2i'+1)}\left\{
\begin{array}{ccc}
I_c & i & I\\
I_a & I'& i'
\end{array}\right\}  \langle(\lambda_c\mu_c)Y_cI_c;(10)yi||(\lambda\mu)YI\rangle \nonumber \\ & & \times \langle(10)yi;(11)Y_aI_a||(10)y'i'\rangle\;
\langle(\lambda\mu)YI;(11)Y_aI_a||(\lambda'\mu')Y'I'\rangle_{\rho}.
\end{eqnarray}
The required isoscalar factors can be found in Refs. \cite{HECHT,vergados}.

\subsection{Results}

 Our analytic results for the non-vanishing isoscalar factors
\begin{eqnarray}\label{GIA} 
\left(\begin{array}{cc||c}
	[N_c-1,1]    &  [21^4]   & [N_c-1,1]   \\
	(\lambda \mu) S  &  (\lambda^a\mu^a) S^i  &  (\lambda' \mu') S'
      \end{array}\right)_{\rho} , 
   \end{eqnarray}
associated
to the matrix elements of $G^{ia}$ as defined in Eq. (\ref{GEN}) together
with the normalization (\ref{normes})
are exhibited in 
Tables  \ref{octet_spin_one_half}, 
\ref{octet_spin_three_halfs}, \ref{decuplet_spin_one_half} and 
\ref{singlet_spin_one_half}.  For $N_c$ = 3 they correspond to $^28$,  $^48$,
$^2{10}$ and $^2{1}$ multiplets respectively. To make the
applications easier they are expressed in 
terms of $N_c$ and the spin of the total system which is fixed for
each multiplet, namely 1/2, 3/2, 1/2 and 1/2 respectively. The values of the
spin are consistent with the label of the corresponding SU(3)
irreducible representation $(\lambda\mu)$ as illustrated by the examples
(\ref{decupletch52})-(\ref{singletch52}), \emph{i.e.} one has $\lambda=2S$ and $\mu=(N_c-2S)/2$ for all the multiplets.
The index $\rho$ is
specified whenever necessary,  with its two distinct values 1 and 2.

\begin{sidewaystable}
{\scriptsize
 \renewcommand{\arraystretch}{2.5}
\begin{tabular}{l|c|c|l}
\hline
\hline
$(\lambda_1\mu_1)S_1$ \hspace{0.5cm} & \hspace{0.5cm}$(\lambda_2\mu_2)S_2$ \hspace{0.5cm} & \hspace{0.5cm}$\rho$\hspace{0.5cm} & \hspace{0.5cm}$\left(\begin{array}{cc||c}                                         [N_c-1,1]  &  [21^4]  &  [N_c-1,1] \\
                           (\lambda_1\mu_1)S_1 & (\lambda_2\mu_2)S_2 & (\lambda\mu)S
                                      \end{array}\right)_\rho$  \\
\vspace{-0.8cm} &  &   & \\
\hline
$(\lambda\mu)S$\hspace{0.5cm} & \hspace{0.cm}$(11)1$ & $1$ &\hspace{0.5cm}$\left\{12S(S+1)+N_c[4S(S+1)-3]\right\}\sqrt{\frac{2}{S(S+1)\left[N_c(N_c+6)+12S(S+1)\right]N_c(5N_c+18)}}$\\
$(\lambda\mu)S$\hspace{0.5cm} & \hspace{0.cm}$(11)1$ & $2$ &\hspace{0.5cm}$\frac{4S^2(S+1)^2-2N_cS(S+1)-(S^2+S-1)N_c^2}{2S(S+1)}\sqrt{\frac{6(N_c-2S+4)(N_c+2S+6)}{(N_c-2S)(N_c+2S+2)\left[N_c(N_c+6)+12S(S+1)\right]N_c(5N_c+18)}}$\\
$(\lambda\mu)S+1$\hspace{0.5cm} & \hspace{0.cm}$(11)1$ & $1$ &\hspace{0.5cm}$-\frac{3\sqrt{2S(2S+3)(N_c+2S+2)}}{\sqrt{(S+1)(2S+1)\left[N_c(N_c+6)+12S(S+1)\right](5N_c+18)}}$\\
$(\lambda\mu)S+1$\hspace{0.5cm} & \hspace{0.cm}$(11)1$ & $2$ &\hspace{0.5cm}$\frac{N_c}{S+1}\sqrt{\frac{3(2S+3)(N_c-2S+4)(N_c+2S+6)}{2(2S+1)(N_c-2S)\left[N_c(N_c+6)+12S(S+1)\right](5N_c+18)}}$\\
$(\lambda+2,\mu-1)S$\hspace{0.5cm} & \hspace{0.cm}$(11)1$ & $/$ &\hspace{0.5cm}$\frac{1}{S+1}\sqrt{\frac{3(2S+3)(N_c+2S+2)(N_c+2S+6)}{2(2S+1)(N_c+2S+4)(5N_c+18)}}$ \\
$(\lambda-1,\mu-1)S$\hspace{0.5cm} & \hspace{0.cm}$(11)1$ & $/$ &\hspace{0.5cm}$\sqrt{\frac{12(N_c+2S)}{S(2S+1)(N_c-2S+2)(N_c+2S+2)(5N_c+18)}}$\\
$(\lambda\mu)S$\hspace{0.5cm} & \hspace{0.cm}$(11)0$ & $1$ &\hspace{0.5cm}$\sqrt{\frac{N_c(N_c+6)+12S(S+1)}{2N_c(5N_c+18)}}$\\
$(\lambda\mu)S$\hspace{0.5cm} & \hspace{0.cm}$(11)0$ & $2$ &\hspace{0.5cm} 0\\
$(\lambda\mu)S$\hspace{0.5cm} & \hspace{0.cm}$(00)1$ & $/$ &\hspace{0.5cm} $\sqrt{\frac{4S(S+1)}{N_c(5N_c+18)}}$\\ 
\hline
\hline
\end{tabular}}
\caption{Isoscalar factors of the SU(6) generators,
 Eqs. (\ref{normes}) and (\ref{GEN}),
corresponding to 
the $^28$ multiplet.}
\label{octet_spin_one_half} 
\end{sidewaystable}

\begin{sidewaystable}
{\scriptsize
 \renewcommand{\arraystretch}{2.5}
\begin{tabular}{l|c|c|l}
\hline
\hline
$(\lambda_1\mu_1)S_1$ \hspace{0.5cm} & \hspace{0.5cm}$(\lambda_2\mu_2)S_2$ \hspace{0.5cm} & \hspace{0.5cm}$\rho$\hspace{0.5cm} & \hspace{0.5cm}$\left(\begin{array}{cc||c}                                         [N_c-1,1]  &  [21^4]  &  [N_c-1,1] \\
                           (\lambda_1\mu_1)S_1 & (\lambda_2\mu_2)S_2 & (\lambda-2,\mu+1)S
                                      \end{array}\right)_\rho$  \\
\vspace{-0.8cm} &  &   & \\
\hline
$(\lambda-2,\mu+1)S$\hspace{0.5cm} & \hspace{0.cm}$(11)1$ & $1$ &\hspace{0.5cm}$ \left[N_c(4S-3)+6S\right]\sqrt{\frac{2(S+1)}{S\left[N_c(N_c+6)+12(S-1)S\right]N_c(5N_c+18)}}$\\
$(\lambda-2,\mu+1)S$\hspace{0.5cm} & \hspace{0.cm}$(11)1$ & $2$ &\hspace{0.5cm}$-\frac{N_c-2S}{S}\sqrt{\frac{3(S-1)(S+1)(N_c-2S+6)(N_c+2S)(N_c+2S+4)}{2(N_c-2S+2)\left[N_c(N_c+6)+12(S-1)S\right]N_c(5N_c+18)}}$\\
$(\lambda\mu)S-1$\hspace{0.5cm} & \hspace{0.cm}$(11)1$ & $/$ &\hspace{0.5cm}$\frac{N_c+4S^2}{S}\sqrt{\frac{3(N_c+2S+4)}{2(2S-1)(2S+1)(N_c+2S+2)N_c(5N_c+18)}}$\\
$(\lambda-2,\mu+1)S-1$\hspace{0.5cm} & \hspace{0.cm}$(11)1$ & $1$ &\hspace{0.5cm}$\frac{3\sqrt{2(S-1)(N_c+2S)}}{\sqrt{S\left[N_c(N_c+6)+12(S-1)S\right](5N_c+18)}}$ \\
$(\lambda-2,\mu+1)S-1$\hspace{0.5cm} & \hspace{0.cm}$(11)1$ & $2$ &\hspace{0.5cm}$-\frac{N_c}{S}\sqrt{\frac{3(N_c-2S+6)(N_c+2S+4)}{2(N_c-2S+2)\left[N_c(N_c+6)+12(S-1)S\right](5N_c+18)}}$\\
$(\lambda-3,\mu)S-1$\hspace{0.5cm} & \hspace{0.cm}$(11)1$ & $/$ &\hspace{0.5cm}$-2\sqrt{\frac{3(S-1)(N_c+2S-2)}{(2S-1)(N_c-2S+4)N_c(5N_c+18)}}$\\
$(\lambda-2,\mu+1)S$\hspace{0.5cm} & \hspace{0.cm}$(11)0$ & $1$ &\hspace{0.5cm}$\sqrt{\frac{N_c(N_c+6)+12(S-1)S}{2N_c(5N_c+18)}}$\\
$(\lambda-2,\mu+1)S$\hspace{0.5cm} & \hspace{0.cm}$(11)0$ & $2$ &\hspace{0.5cm}$0$\\
$(\lambda-2,\mu+1)S$\hspace{0.5cm} & \hspace{0.cm}$(00)1$ & $/$ &\hspace{0.5cm}$\sqrt{\frac{4S(S+1)}{N_c(5N_c+18)}}$\\
\hline
\hline
\end{tabular}}
\caption{Isoscalar factors of the SU(6) generators, Eqs. (\ref{normes}) and (\ref{GEN}),
corresponding to 
the $^48$ multiplet.}
\label{octet_spin_three_halfs} 
\end{sidewaystable}

\begin{sidewaystable}
{\scriptsize
 \renewcommand{\arraystretch}{2.5}
\begin{tabular}{l|c|c|l}
\hline
\hline
$(\lambda_1\mu_1)S_1$ \hspace{0.5cm} & \hspace{0.5cm}$(\lambda_2\mu_2)S_2$ \hspace{0.5cm} & \hspace{0.5cm}$\rho$\hspace{0.5cm} & \hspace{0.5cm}$\left(\begin{array}{cc||c}                                         [N_c-1,1]  &  [21^4]  &  [N_c-1,1] \\
                           (\lambda_1\mu_1)S_1 & (\lambda_2\mu_2)S_2 & (\lambda+2,\mu-1)S
                                      \end{array}\right)_\rho$  \\
\vspace{-0.8cm} &  &   & \\
\hline
$(\lambda+2,\mu-1)S$\hspace{0.5cm} & \hspace{0.cm}$(11)1$ & $1$ &\hspace{0.5cm}$\left[N_c(4S+7)+6(S+1)\right]\sqrt{\frac{2S}{(S+1)\left[N_c(N_c+6)+12(S+1)(S+2)\right]N_c(5N_c+18)}}$\\
$(\lambda+2,\mu-1)S$\hspace{0.5cm} & \hspace{0.cm}$(11)1$ & $2$ &\hspace{0.5cm}$-\frac{N_c+2(S+1)}{S+1}\sqrt{\frac{3S(S+2)(N_c-2S-2)(N_c-2S+2)(N_c+2S+8)}{2(N_c+2S+4)\left[N_c(N_c+6)+12(S+1)(S+2)\right]N_c(5N_c+18)}}$\\
$(\lambda\mu)S+1$\hspace{0.5cm} & \hspace{0.cm}$(11)1$ & $/$ &\hspace{0.5cm}$\frac{N_c+4(S+1)^2}{S+1}\sqrt{\frac{3(N_c-2S+2)}{2(2S+1)(2S+3)(N_c-2S)N_c(5N_c+18)}}$ \\
$(\lambda\mu)S$\hspace{0.5cm} & \hspace{0.cm}$(11)1$ & $/$ &\hspace{0.5cm} $-\frac{1}{S+1}\sqrt{\frac{3(N_c+2S+2)(N_c-2S+2)}{2(N_c-2S)(5N_c+18)}}$\\
$(\lambda+2,\mu-1)S$\hspace{0.5cm} & \hspace{0.cm}$(11)0$ & $1$ &\hspace{0.5cm}$\sqrt{\frac{N_c(N_c+6)+12(S+1)(S+2)}{2N_c(5N_c+18)}}$\\
$(\lambda+2,\mu-1)S$\hspace{0.5cm} & \hspace{0.cm}$(11)0$ & $2$ &\hspace{0.5cm} 0\\
$(\lambda+2,\mu-1)S$\hspace{0.5cm} & \hspace{0.cm}$(00)1$ & $/$ &\hspace{0.5cm} $\sqrt{\frac{4S(S+1)}{N_c(5N_c+18)}}$\\
\hline
\hline
\end{tabular}}

\caption{Isoscalar factors of the SU(6) generators, Eqs. (\ref{normes}) and (\ref{GEN}) ,
corresponding to 
the $^210$ multiplet.}\label{decuplet_spin_one_half} 
\end{sidewaystable}

\begin{sidewaystable}
{\scriptsize
 \renewcommand{\arraystretch}{2.5}
\begin{tabular}{l|c|c|l}
\hline
\hline
$(\lambda_1\mu_1)S_1$ \hspace{0.5cm} & \hspace{0.5cm}$(\lambda_2\mu_2)S_2$ \hspace{0.5cm} & \hspace{0.5cm}$\rho$\hspace{0.5cm} & \hspace{0.5cm}$\left(\begin{array}{cc||c}                                         [N_c-1,1]  &  [21^4]  &  [N_c-1,1] \\
                           (\lambda_1\mu_1)S_1 & (\lambda_2\mu_2)S_2 & (\lambda-1,\mu-1)S
                                      \end{array}\right)_\rho$  \\
\vspace{-0.8cm} &  &   & \\
\hline
$(\lambda-1,\mu-1)S$\hspace{0.5cm} & \hspace{0.cm}$(11)1$ & $1$ &\hspace{0.5cm}$\left[N_c(4S-3)+6S\right]\sqrt{\frac{2(S+1)}{S\left[N_c^2+12(S^2-1)\right]N_c(5N_c+18)}}$\\
$(\lambda-1,\mu-1)S$\hspace{0.5cm} & \hspace{0.cm}$(11)1$ & $2$ &\hspace{0.5cm}$-\left\{N_c(N_c+6)-4\left[S(S-1)-3\right]\right\}\sqrt{\frac{3(2S-1)(S+1)(N_c-2S-2)(N_c+2S-2)}{2S(2S+1)(N_c-2S+2)(N_c+2S+2)\left[N_c^2+12(S^2-1)\right]N_c(5N_c+18)}}$\\
$(\lambda\mu)S+1$\hspace{0.5cm} & \hspace{0.cm}$(11)1$ & $/$ &\hspace{0.5cm}$\sqrt{\frac{6(2S+3)(N_c+2S+4)}{(2S+1)(N_c-2S)N_c(5N_c+18)}}$\\
$(\lambda\mu)S$\hspace{0.5cm} & \hspace{0.cm}$(11)1$ & $/$ &\hspace{0.5cm}$\frac{1}{S}\sqrt{\frac{6(N_c+2S+4)}{(N_c-2S)(N_c+2S+2)(5N_c+18)}}$\\
$(\lambda-1,\mu-1)S$\hspace{0.5cm} & \hspace{0.cm}$(11)0$ & $1$ &\hspace{0.5cm}$\sqrt{\frac{N_c^2+12(S^2-1)}{2N_c(5N_c+18)}}$\\
$(\lambda-1,\mu-1)S$\hspace{0.5cm} & \hspace{0.cm}$(11)0$ & $2$ &\hspace{0.5cm} 0 \\
$(\lambda-1,\mu-1)S$\hspace{0.5cm} & \hspace{0.cm}$(00)1$ & $/$ &\hspace{0.5cm} $\sqrt{\frac{4S(S+1)}{N_c(5N_c+18)}}$\\ 
\hline
\hline
\end{tabular}}
\caption{Isoscalar factors of the SU(6) generators, Eqs. (\ref{normes}) and (\ref{GEN}), 
corresponding to  the $^21$ multiplet.}
\label{singlet_spin_one_half} 
\end{sidewaystable}

\section{The mass operator of strange and non-strange baryons}

It is very important to apply the $1/N_c$ expansion
method to both non-strange and strange baryons together.
First, we have at our disposal a larger number of experimental
data than for non-strange baryons alone 
and second, we can get a unified picture of all light baryons.
 
When the SU(3)-flavor symmetry is exact, the $1/N_c$ expansion 
mass operator describing an excited state 
can be written as the linear combination 
\begin{equation}
\label{massoperator1}
M^{(1)} = \sum_{i} c_i O_i ,
\end{equation} 
where $c_i$ are unknown coefficients which parametrize the QCD dynamics
and the operators $O_i$ are of type 
\begin{equation}\label{OLFS}
O_i = \frac{1}{N^{n-1}_c} O^{(k)}_{\ell} \cdot O^{(k)}_{SF}
\end{equation}
where  $O^{(k)}_{\ell}$ is a $k$-rank tensor in SO(3) and  $O^{(k)}_{SF}$
a $k$-rank tensor in SU(2), but scalar in SU(3)-flavor (as shown by the 
upper index of $M^{(1)}$).
This implies that $O_i$ is a combination 
of SO(3) generators $L^i$ and of SU(6) generators. The presence 
of $L^i$ is necessary in describing excited states.

When the SU(3)-flavor symmetry is broken the mass operator in the $1/N_c$
expansion contains additional terms,
as first performed in Ref.
\cite{Jenkins:1995td} for the symmetric baryon multiplet
\begin{equation}
\label{massoperator}
M = \sum_{i} c_i O_i + \sum_{i} d_i B_i,
\end{equation} 
where the operators $B_i$ are defined to have zero expectation
values for nonstrange quarks. 
The values of the coefficients $c_i$  and $d_i$ are found by  
a numerical fit to data.

\begin{figure}[h!]
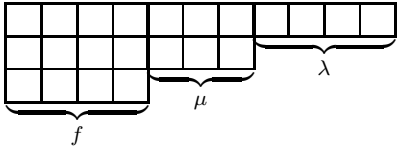

$$\underbrace{\begin{Young} & & & \cr & & & \cr & & & \cr \end{Young}}_{f}\negthinspace\raisebox{12.5pt}{$\underbrace{\begin{Young}
    & & \cr
    & & \cr
   \end{Young}}_{\mu}$}\ \negthinspace
   \negthinspace\raisebox{25pt}{$\underbrace{\begin{Young} & & & \cr
   \end{Young}}_{\lambda}$}$$ 
\caption{Young diagram of an SU(3) irreducible representation.}
\label{su3young}
\end{figure}

An essential step is to find all linearly independent operators 
contributing to a given order $\mathcal{O}(1/N_c)$. In Table \ref{operators} 
we present a list of operators expected to be dominant up to order $1/N_c$.
The order of their matrix elements in SU(6), indicated in the
second column of  the table
is not always the same as
in SU(4) \cite{Matagne:2006dj}.
For example the operator $O_4$ is of order $N^0_c$ while in SU(4) 
is of order $1/N_c$. This can be understood by looking at  
its matrix elements obtained from the relations (\ref{FLAVOR})-(\ref{CSU3})

\begin{equation}\label{O4}
 \frac{1}{N_c}T^aT^a = 
 \frac{1}{12N_c}\left\{N_c(N_c+6)+3\lambda(\lambda+2)-3f[2(N_c+3)-3f]\right\},
\label{ttsu3}
\end{equation}
with $\lambda$, $\mu$  and $f$ illustrated by the example in Figure \ref{su3young}. 
Let us remind that 
for any irreducible representation of SU(3) two labels are enough.
Usually one takes 
$\lambda$ and $\mu$. For the present discussion, where $N_c$ is needed,
it is more convenient to
use $\lambda$ and $f$.  
For a system of $N_c$ quarks  one has
$\mu=(N_c-\lambda-3f)/2$,  
which leads to Eq. (\ref{O4}). When $N_c$ = 3 one has $f=0$ for the octet  
and the decuplet and $f=1$ for the singlet of SU(3).  By looking at 
Eq. (\ref{ttsu3}), one can notice that the first term in the bracket,
$N_c(N_c+6)$,
which is responsible for the order of the operator $O_4$, 
is the same for all representations. That justifies the 
new definition introduced in Table \ref{operators} where 
$(N_c+6)/12$ has been subtracted. 
It is important to stress that this new definition
of $O_4$ gives  the same matrix elements   
as the isospin-isospin operator $1/N_c( T^aT^a)$,
used in Ref. \cite{Matagne:2006dj} for non-strange baryons
($\lambda=2I$ for non-strange baryons). 
The new operator $O_4$ is then a natural generalization to SU(3) of 
its SU(2)-isospin counterpart.

An important property 
is that the order of $O_4$ is now $N_c^0$ 
and not  $1/N_c$ as one would expect from SU(4). This comes from the third 
term of 
Eq. (\ref{ttsu3})
which contributes only for 
SU(6) representations which become singlets for $N_c = 3$, as explained above. 
The SU(6) symmetry is then broken to order 1 in the large $N_c$ 
limit for the mixed symmetric representation $[N_c-1,1]$.  
This result, which does not appear for ground state baryons, 
should be analyzed in more details in the future.
Meanwhile one can notice that the order $N_c^0$ of $O_4$
is consistent with Eqs. (18) and (19) of  Ref. \cite{Cohen:2005ct}
where one predicts five towers of states and where the singlet  
always belongs to different towers from the octet and the decuplet.
By adding  more operators in the restricted list of Table \ref{operators} 
as, for example, $L^{(2)ij} G^{ia} G^{ja}$, we might expect to obtain results
consistent with Ref. \cite{Cohen:2005ct} in physical applications.
 
\begin{table*}[h!]
\label{operators}{\scriptsize
\renewcommand{\arraystretch}{2} 
\begin{tabular}{lr}
\hline
\hline
Operator \hspace{2cm} &\hspace{0.0cm} Matrix elements order in SU(6)\\
\hline

$O_1 = N_c \ \1 $                               &  $N_c$      \\
$O_2 = L^i S^i$                	        &  $N^0_c$    \\
$O_3 = \frac{1}{N_c}S^iS^i$                     &  $N^{-1}_c$ \\
$O_4 = \frac{1}{N_c}\left(T^aT^a-\frac{1}{12}N_c(N_c+6)\right)$                     &  $N^{0}_c$ \\
$O_5 = \frac{3}{N_c} L^iT^aG^{ia}$            &  $N^0_c$    \\
$O_6 =  \frac{3}{N_c^2} S^i T^a G^{ia}$         &  $N^{-1}_c$  \\
$O_7 =  \frac{1}{N_c} L^{(2)ij}S^iS^i$         &  $N^{-1}_c$ \\
\hline
$B_1 = \mathcal{S}$  &  $N^0_c$  \\
\hline \hline
\end{tabular}}
\caption{Examples of operators $O_i$ and $B_1$ entering the mass formula.}
\end{table*}


The order of $O_5$ and $O_6$ follow from the arguments given in Ref. \cite{JENKINS}.
Accordingly, unlike the case of two light flavors, the matrix elements 
of the flavor generators $T^a$ and spin-flavor generators $G^{ia}$ do not
have the same $N_c$ dependence everywhere in the flavor weight
diagram. 
The baryons under concern, located at the top of an SU(3) weight diagram,
therefore having finite strangeness, have matrix elements of  
$T^a$ which are of order $\mathcal{O}(1)$,
$\mathcal{O}(\sqrt{N_c})$ and $\mathcal{O}(N_c)$ for
$a = 1,2,3,$  $a = 4,5,6,7$ and $a = 8$ respectively
and matrix elements of $G^{ia}$ which are $\mathcal{O}(N_c)$, 
$\mathcal{O}(\sqrt{N_c})$   
and $\mathcal{O}(1)$. Then the corresponding combinations give
for $O_5$ and $O_6$  the order $N^0_c$ and $N_c^{-1}$ respectively. $B_1$ is 
of course of order $N_c^0$. 

The contribution of SO(3), is first manifested in the 
operator $O_2$ which represents the spin-orbit coupling where $L^i$ and $S^i$ 
are the total angular momentum and spin components. In applications 
one can also use the Hartree approximation \cite{CCGL} 
which is a one-body operator. This  approximation is useful because it shows
that the leading order of the spin-orbit contribution is $N^0_c$.
Lastly, the operator $O_7$ contains the SO(3) 2nd rank tensor, defined as
\begin{equation}\label{TENSOR}
L^{(2)ij} = \frac{1}{2}\left\{L^i,L^j\right\}-\frac{1}{3}
\delta_{i,-j}\vec{L}\cdot\vec{L},
\end{equation}
which, like $L^i$, acts on the orbital wave function $|L m_{L} \rangle$  
of the whole system of $N_c$ quarks. 

In SU(4) the practice on the $[{\bf 70},1^-]$ multiplet \cite{Matagne:2006dj}
showed that the operators $O_1$, $O_3$, $O_4$ (defined as
$  \frac{1}{N_c}T^aT^a $)
and $O_6$ are the most
dominant. 
A problem is to find out if this behavior 
also holds in SU(6). Also one has to reanalyze
the $[{\bf 70},\ell^+]$ 
multiplet, studied so far in the decoupling scheme 
only \cite{Matagne:2006zf}.
Applications to baryons belonging to the $[{\bf 70},\ell^\pm]$
multiplets  will be considered in subsequent studies.

We should finally mention, that the method based on the separation
of the system into a symmetric core of $N_c-1$ quarks and an
excited quark, as first used in Ref.\ \cite{CCGL}, acquired some
support from the work of Pirjol and Schat \ \cite{Pirjol:2007ed}, 
based on a large $N_c$ quark model Hamiltonian, where explicit 
results for the coefficients $c_i$ are presented both for
the one gluon exchange (OGE)  and for the Goldstone boson exchange 
(GBE) hyperfine interactions, with radial dependent form factors.
An extension of the study of Collins and Georgi  \cite{Collins:1998ny} 
from $N_c = 3$ to large $N_c$ is obtained in this way.  

Including the space degree of freedom, 
Pirjol and Schat decompose the two-body operators into tensor operators 
transforming  as S, MS and E representations of $S_{N_c}$
(the latter exists only for $N_c > 3$). Moreover, the splitting 
of the SU(4) generators into two pieces (see introduction), 
$S^i = S^i_c+s^i$, $T^a = T^a_c+t^a$ and $G^{ia} = G^{ia}_c+ g^{ia}$,
matches the choice of their basis states.
The conclusion of Ref. \cite{Pirjol:2007ed} was  that the inclusion of core 
and excited quark operators
is necessary, at variance with our simplified procedure. 

A useful result is that  Pirjol and Schat obtain 
a large $N_c$ dependence similar to that of the $1/N_c$ expansion method. 
The tower structure, 
first observed in the N = 1 band in Ref.  \cite{Pirjol:1997sr}, is satisfied
at leading order in $1/N_c$. The resulting mass 
formula  contains 6 independent non-vanishing coefficients
at order $1/N_c$ both  for OGE and for GBE.
They have to be found by fitting the 
7 resonance masses available in the  N = 1 band.
Therefore some arbitrariness is 
imposed in combining OGE with GBE.  In general, the arbitrariness 
could lead to anomalies, as shown in Ref. \cite{Matagne:2008fw}.

On the other hand in the method based on the quark model Hamiltonian,
the operator  $S^iT^aG^{ia}$ is absent, but present in our case,
see Table \ref{operators}. 
In Ref. \cite{Matagne:2006dj}
we have shown that such an operator brings a dominant contribution to
most of the nucleon masses in the N = 1 band.
According to Ref.  \cite{CCGL} this absence is allowed in SU(4) 
but, as explicitly stated  there, 
for more than two flavors a term like $S^i_ct^aG^{ia}_c$,
should be included in the mass operator. 
Thus the extension to SU(6) of the work of Pirjol and Schat could, 
at least in this respect, raise problems.

The basic difference between our work 
and that of Pirjol and Schat is
the radial dependence is integrated out in our case, 
consistent with the $1/N_c$ expansion method.  
When there is no radial dependence,  the spin-spin operator is symmetric.
The quadrupole operator of the quark model makes its presence 
through its symmetric part, which is related to the 2nd-rank SO(3) operator 
(\ref{TENSOR}), as in the work of  Collins and Georgi \cite{Collins:1998ny}, 
or of Carlson et al.  \cite{CCGL} and as in subsequent studies.
So far, in applications,  the spin-orbit term was treated in  
the Hartree approximation.  In this context, our  
operator basis is complete inasmuch as the orbital part is always symmetric, 
thus the flavour-spin part should be symmetric too, so we have only
$\langle \mathrm{MS} | O^S | \mathrm{MS} \rangle$ matrix elements 
in the flavour-spin space.
  
The differences between the two methods should be confronted in
future applications to nonstrange and strange baryons, from where 
one wishes to obtain meaningful physical information on the 
coefficients $c_i$.

\section{Conclusions}
The isoscalar factors of the SU(6) generators derived in this study opens
new applications of the $1/N_c$ expansion method to baryons spectroscopy.
It allows to combine data on non-strange and strange baryons together
and to lead to a more precise determination of the coefficients
$c_i$ and $d_i$ which encode the QCD dynamics.  
Finally, our results presented in Tables I-IV,  
can be used for other $N$-body problems governed by SU(6) symmetry and
where the spin is known.


\end{document}